\documentclass[a4paper,fleqn,usenatbib]{mnras}
\usepackage{newtxtext}
\usepackage{mathptmx}
\usepackage[T1]{fontenc}
\usepackage{ae,aecompl}
\usepackage{graphicx}	
\usepackage{amsmath}	
\usepackage{amssymb}	
\usepackage{color}

\title[GLADE: A Galaxy List for the Advanced Detector Era]{GLADE: A Galaxy Catalogue for Multi-Messenger Searches in the Advanced Gravitational-Wave Detector Era}
\author[D\'alya, Galg\'oczi, Dobos, Frei, Heng, Macas, Messenger, Raffai \& de Souza]{G. D\'alya$^{1,2}$\thanks{E-mail: dalyag@caesar.elte.hu (DG);}, G. Galg\'oczi$^{1,2}$, L. Dobos$^{1}$, Z. Frei$^{1,2}$, I. S. Heng$^{3}$, R. Macas$^{4}$, \newauthor C. Messenger$^{3}$, P. Raffai$^{1,2}$ and R. S. de Souza$^{5}$ \\
	$^1$Institute of Physics, E\"otv\"os University, 1117 Budapest, Hungary \\
	$^2$MTA-ELTE Astrophysics Research Group, 1117 Budapest, Hungary\\
    $^3$SUPA, University of Glasgow, Glasgow G12 8QQ, United Kingdom\\
    $^4$School of Physics and Astronomy, Cardiff University, Cardiff CF24 3AA, United Kingdom\\
    $^5$Department of Physics \& Astronomy, University of North Carolina at Chapel Hill, Chapel Hill, NC 27599-3255, USA\\
         }

\date{Accepted XXX. Received YYY; in original form ZZZ}
\pubyear{2018}

\begin{document}
\label{firstpage}
\pagerange{\pageref{firstpage}--\pageref{lastpage}}
\maketitle

\begin{abstract}
We introduce a value-added full-sky catalogue of galaxies, named as Galaxy List for the Advanced Detector Era, or GLADE. The purpose of this catalogue is to (i) help identifications of host candidates for gravitational-wave events, (ii) support target selections for electromagnetic follow-up observations of gravitational-wave candidates, (iii) provide input data on the matter distribution of the local universe for astrophysical or cosmological simulations, and (iv) help identifications of host candidates for poorly localised electromagnetic transients, such as gamma-ray bursts observed with the InterPlanetary Network. Both being potential hosts of astrophysical sources of gravitational waves, GLADE includes inactive and active galaxies as well. GLADE was constructed by cross-matching and combining data from five separate (but not independent) astronomical catalogues: GWGC, 2MPZ, 2MASS XSC, HyperLEDA and SDSS-DR12Q. GLADE is complete up to $d_L=37^{+3}_{-4}\ \mathrm{Mpc}$ in terms of the cumulative $B$-band luminosity of galaxies within luminosity distance $d_L$, and contains all of the brightest galaxies giving half of the total $B$-band luminosity up to $d_L=91\ \mathrm{Mpc}$. As $B$-band luminosity is expected to be a tracer of binary neutron star mergers (currently the prime targets of joint GW+EM detections), our completeness measures can be used as estimations of completeness for containing all binary neutron star merger hosts in the local universe.
\end{abstract}

\begin{keywords}
catalogues --- galaxies: distances and redshifts.
\end{keywords}

\section{Introduction}
\label{sec:Introduction}
Advanced LIGO (aLIGO, see \citealt{2015CQGra..32g4001L}) and Advanced Virgo (AdV, see \citealt{2015CQGra..32b4001A}) are second-generation gravitational-wave (GW) detectors located in Hanford, US-WA, Livingston, US-LA, and near Cascina, Italy, respectively. The two aLIGO detectors began their second observing run (O2) on 30 November 2016, with an improved network sensitivity compared to the first observing run (O1, during which the first detections of GWs from binary black hole mergers were accomplished, see \citealt{2016PhRvX...6d1015A}). AdV joined the network of aLIGO detectors during O2, on 1 August 2017 \citep{2017PhRvL.119n1101A}.

An ongoing effort of the emergent field of multi-messenger astronomy (see e.g. \citealt{2011GReGr..43..437C})  is to discover electromagnetic (EM) counterparts of transient GW events with targeted EM follow-up observations \citep{2016LRR....19....1A}. Maps of posterior probability densities for sky positions of GW source candidates are being distributed by the LIGO-Virgo Collaboration to astronomer partners, who then carry out follow-up observations at various EM wavelengths from radio to gamma bands (see e.g. \citealt{2017ApJ...848L..12A}, \citealt{2016PhRvD..94l2007A}, \citealt{2016ApJ...826L..13A}, \citealt{2016ApJ...826L..29C}).

The 90 percent credible localisation areas of transient GW sources produced with the two aLIGO detectors currently cover hundreds of square degrees \citep{2016LRR....19....1A} (the exact size depends on the source location relative to the detectors, and the signal-to-noise ratio of the signal, see e.g. \citealt{2017ApJ...839...15B}, \citealt{2015ApJ...804..114B}, and \citealt{2015ApJ...800...81E}), which is at least an order of magnitude larger than the field of view of most of the EM follow-up instruments. This makes the identification of EM counterparts challenging even with wide-field telescopes, due to the large number of pointings and long integration times required for follow-ups, as well as due to the false positives potentially obtained from these observations. Because of this, the optimization of EM follow-up observing strategies is an important task and a widely researched topic, discussed in details in e.g. \cite{2012A&A...541A.155A}, \cite{2014ApJS..211....7A}, \cite{2014ApJ...795..105S}, \cite{2016ApJ...820..136G} and \cite{2018arXiv180302255C}. Since in the upcoming years more GW detectors, such as KAGRA \citep{2012CQGra..29l4007S} and LIGO-India \citep{LIGO_India}, are coming online, GW source localisations are expected to greatly improve. For example, it is estimated that more than 20 percent of the 90 percent credible localisation areas will have sizes smaller than 5 square degrees when produced by the 4-detector network of aLIGO, AdV, and LIGO-India in 2024 \citep{2016LRR....19....1A}.

Joint detections of temporally coincident GWs and high-energy neutrinos (HEN) would also present multiple advantages compared to the detection of GWs or HENs only, such as increased search sensitivity, or introducing constrains on the population of astrophysical GW+HEN sources \citep{2012PhRvD..85j3004B}. The distribution of potential sources, inferred from galaxy catalogues, can be used in joint GW+HEN searches to increase sensitivity and reject false coincidences.

Binary neutron star (BNS) coalescences followed by possible kilonova emissions (see e.g. \citealt{1998ApJ...507L..59L}, \citealt{2010MNRAS.406.2650M}, \citealt{2013Natur.500..547T}) are currently the most promising sources for joint GW and EM observations. The network of aLIGO and AdV detectors achieved the detection of such an event on August 17, 2017 \citep{2017PhRvL.119p1101A}, which lead to the discovery of the EM counterpart in various EM bands and to the identification of the host galaxy \citep{2017ApJ...848L..12A}, partially with the help of the catalogue presented here (see Section \ref{sec:Applications} for details). Kilonovae have observable EM emissions lasting for $\gtrsim$~1 week in near infrared. This is a reasonable time frame after a GW detection to search through galaxies from an existing catalogue for a fading EM counterpart, but can be too short to cover the whole localisation area with EM observations deep enough to achieve the detection of the EM counterpart. Furthermore, extending such an existing galaxy catalogue with rapid galaxy surveys within the GW localisation areas, carried out by dedicated small telescopes, is also feasible in this timescale \citep{2015ApJ...801L...1B}. 

As it has been shown in multiple papers (e.g. \citealt{2012A&A...539A.124L}, \citealt{2014ApJ...784....8H}), using galaxy catalogues in target selections for follow-up observations greatly increases the chance of detecting the EM counterpart, even if the catalogue is incomplete. Since we expect GW sources to reside in or near galaxies, restricting follow-up observations to galaxies can decrease significantly the required number of pointings, the total integration time, and the number of false positives found. Galaxies inside the localisation volume (see e.g. \citealt{2018arXiv180108009D} and \citealt{2016ApJ...829L..15S}) of a GW event can be ranked in terms of probability of hosting a BNS detected by GW detectors, based on e.g. their $B$-band luminosities (see \citealt{2014ApJ...784....8H} for details), and astronomers can observe these host candidates in a sequential order from high to low probabilities, thereby enhancing the likelihood for EM counterpart detection within a more limited time.

Alternative applications of a galaxy catalogue include identifying host candidates for poorly localised EM transients, see e.g. \cite{2014ApJ...795...43F}. For example, there have been several searches for GW signals associated with gamma-ray bursts (GRBs) using data from the LIGO and Virgo detectors (see e.g. \citealt{2014PhRvL.113a1102A}). By using the time and sky location information from EM observations, the parameter space of the GW search can be significantly reduced, which improves the efficiency of the GW search. A galaxy catalogue can be used to identify potential host galaxies for poorly localised (e.g. by the InterPlanetary Network or IPN, see \citealt{2013EAS....61..459H}) GRBs, that can be of particular interest for the GW community if the GRB error box or its close proximity contains galaxies that reside within the horizon of advanced GW detectors \citep{2016arXiv161107947L}. Using the results from the GW search and the distance information of the host candidates, we can derive implications on source models, potentially leading to rejecting some of the host candidates or even some of the source models. A galaxy catalogue with high completeness can be useful for astrophysical and cosmological simulations as well, providing prior information on the matter distribution of the local universe (see e.g. \citealt{1978AJ.....83..845S}). Galaxy catalogues will be important for statistical cosmological inference as suggested by \cite{1986Natur.323..310S} and \cite{2012PhRvD..86d3011D}.

As models suggest that the number of possible sources of coincident GW and EM emission in a galaxy is traced by the total $B$-band luminosity of the galaxy (through active star formation, see e.g. \citealt{1991ApJ...380L..17P}), a galaxy catalogue supporting follow-up searches should be as complete as possible in terms of the cumulative $B$-band luminosity of its galaxies \citep{2012PhRvD..85j3004B}. Note however, that according to \cite{2013ApJ...769...56F} a quarter of short GRBs (produced by BNS mergers, see e.g. \citealt{2013Natur.500..547T}) occur in elliptical galaxies, that have low star forming rates, and thus low $B$-band luminosities. This suggests that BNS merger rates for individual galaxies are better estimated by a weighed combination of the galaxies' $B$-band luminosities and stellar masses, where the weighing can depend on the galaxies' morphological types.

In the initial detector era, several EM follow-up partners used the Gravitational Wave Galaxy Catalog (GWGC, see \citealt{2011CQGra..28h5016W}) for observational target selections. This catalogue contains galaxies only within $\sim$100 Mpc, and with the since-updated distance data for its entries GWGC is only complete within $\sim$30 Mpc in terms of cumulative $B$-band luminosity, with the completeness falling rapidly at greater distances (see Section \ref{sec:Completeness} for details). During the initial detector era, the BNS ranges\footnote{The BNS range is the average distance from which a GW detector can detect a circular binary of two 1.4 $M_{\odot}$ neutron stars with a signal-to-noise ratio of 8, where the average is calculated over all possible sky positions and orbital inclinations \citep{2016LRR....19....1A}. Note that the maximum distance from which a BNS (i.e. with optimal sky direction and inclination) can be detected is $\sim2.26$ times larger \citep{1993PhRvD..47.2198F}.} for the individual GW detectors did not exceed $\sim$30 Mpc \citep{2012arXiv1203.2674T}, and thus GWGC served the EM target selections well in this past era. Advanced GW detectors have larger ranges for BNS coalescences, which can be further extended if multiple detectors operate as a coherent network in GW searches. As a result, more distant galaxies, even at $\sim$300 Mpc or further \citep{2016MNRAS.455.1522E} can be potential hosts of detectable GW+EM sources, and therefore should be included in a galaxy catalogue that aims to support EM follow-up efforts.

We have constructed a galaxy catalogue called Galaxy List for the Advanced Detector Era (GLADE) in order to support the previously described EM follow-up efforts, and to meet the challenges imposed on galaxy catalogues by the improved sensitivities of advanced GW detectors. Both being potential hosts of astrophysical sources of GWs, GLADE includes inactive and active galaxies as well (where the fact that a GLADE object is an active galaxy hosting a quasar is indicated in the catalogue). The aim of this paper is to describe the construction and properties of GLADE, as well as to point out the applicability of it in target selections for EM follow-up observations, and in other selected areas. 

This paper is organized as follows. In Section \ref{sec:Compilation}, we introduce methods and catalogues used for constructing GLADE. In Section \ref{sec:Completeness}, we describe the completeness of GLADE with two different methods that use the measured $B$-band luminosities of galaxies and that allows comparisons of the completeness of GLADE with that of other existing full-sky catalogues' (e.g. GWGC, and the Census of the Local Universe or CLU presented in \citealt{2016ApJ...820..136G}). In Section \ref{sec:Applications}, we present how GLADE is already in use for identifying host candidates of GW events by multiple collaborations. Furthermore, we suggest applications of GLADE beyond GW-triggered EM observations, including one where we use GLADE in an automated process to identify potential host galaxies for poorly localised GRBs detected by the IPN. Finally, in Section \ref{sec:Conclusions}, we summarize our conclusions and outline future plans for improving GLADE in terms of completeness and accuracy of its parameters.

Throughout this paper we adopt a flat $\Lambda$CDM cosmology with the following parameters: $H_0 = 100\, h = 70 \ \mathrm{km} \ \mathrm{s}^{-1} \ \mathrm{Mpc}^{-1}$, $\Omega_{\mathrm{M}}=0.27$ and $\Omega_{\Lambda}=0.73$.

\section{Catalogue compilation and statistics}
\label{sec:Compilation}

In the construction of GLADE, we started with cross-matching five separate (but not independent) astronomical catalogues: the GWGC\footnote{\url{vizier.u-strasbg.fr/viz-bin/VizieR?-source=GWGC}} \citep{2011CQGra..28h5016W}, the \mbox{HyperLEDA} catalogue\footnote{\url{leda.univ-lyon1.fr/}} \citep{2014A&A...570A..13M}, the 2 Micron All-Sky Survey Extended Source Catalog\footnote{\url{ipac.caltech.edu/2mass/}}  (2MASS XSC, see \citealt{2006AJ....131.1163S}), the 2MASS Photometric Redshift Catalog\footnote{\url{ssa.roe.ac.uk/TWOMPZ.html}} (2MPZ, see \citealt{2014ApJS..210....9B}), and the Sloan Digital Sky Survey quasar catalogue from the 12th data release\footnote{\url{sdss.org/dr12/algorithms/boss-dr12-quasar-catalog}} (SDSS-DR12Q, see \citealt{Paris}). In this section, we describe the relevant characteristics of these five catalogues, and discuss the cross-matching method we applied and results we obtained.

GWGC is a catalogue of $\sim$50,000 galaxies and $\sim$150 globular clusters, that is a result of merging three existing catalogues: the Tully Nearby Galaxy Catalog \citep{1987ApJ...321..280T}, the Catalog of Neighboring Galaxies \citep{2004AJ....127.2031K}, and the V8k catalogue \citep{2009AJ....138..323T}. In the construction of GWGC, authors of \cite{2011CQGra..28h5016W} used the HyperLEDA catalogue to provide supplemental data (e.g. position angles) for objects where such was available. GWGC does not contain galaxies beyond luminosity distance $d_L\simeq~100~$Mpc due to various recession velocity cuts applied in the catalogues GWGC has been created from. GWGC includes $B$ magnitude and luminosity distance data, along with their errors, for nearly all its entries. Luminosity distances were derived from distance measurements that used a variety of methods, each having had their own measurement errors within the range of 10 to 20 percent. $B$ magnitudes in GWGC have an average error of $\Delta B=0.37^m$. We found that $B$ magnitude values of globular clusters in GWGC were not reliable, therefore we have corrected them using the newest data from the VizieR database\footnote{\url{vizier.u-strasbg.fr/}} (\citealt{2000A&AS..143...23O}).

HyperLEDA is a catalogue of over 3 million objects, created by merging the LEDA (Lyon-Meudon Extragalactic Database, see \citealt{1988ESOC...28..435P}) and Hypercat \citep{1996A&A...309..749P} databases. We only kept $\sim$2.6 million of these objects that are identified as either galaxies or quasars, omitting stars and nebulae. Distance moduli from spectroscopic redshift measurements are given for each entry in HyperLEDA, with a 36 percent corresponding mean error of luminosity distances.

The 2MASS XSC catalogue consists of $\sim$1.6 million objects, each having coordinates and photometric magnitude data in the $J$, $H$ and $K_s$ infrared bands. Sources with apparent magnitude $K_s<13.5^{m}$ 
(i.e. $\sim$50 percent of the sources) are classified as galaxies with 98 percent confidence \citep{2000AJ....119.2498J}, nevertheless we included all 2MASS XSC objects in our cross-matching process, even the ones not satisfying this criterion. Since 2MASS XSC objects are all extended sources beyond the point spread function, and classification tests performed by \cite{2006AJ....131.1163S} implemented filters to exclude double and triple stars, we expect that the stellar contamination of the whole sample is negligible. Since 2MASS is an infrared survey, $B$ magnitudes and redshifts (both of which are important parameters for EM follow-up target selections) are not included in the 2MASS XSC catalogue. Despite the lack of these parameters, this catalogue is an important source of infrared magnitudes, which can be used for estimating stellar masses of galaxies \citep{2003ApJS..149..289B} and thus as an alternative tracer of BNS mergers (see e.g. \citealt{2013ApJ...769...56F} and \citealt{2014ApJ...784....8H}).

Authors of \cite{2014ApJS..210....9B} constructed the 2MPZ catalogue by cross-matching 2MASS XSC with the Wide-field Infrared Survey Explorer catalogue \citep{2010AJ....140.1868W} and the SuperCOSMOS optical catalogue \citep{2001MNRAS.326.1279H}. 2MPZ contains both $B$ magnitudes and photometric redshifts for its more than 900,000 galaxy entries. Bilicki et al. computed photometric redshifts with an artificial neural network algorithm trained on redshift surveys. Errors of these photometric redshifts are not given for the specific entries, but they are nearly independent of distance, and have an all-sky average of  $\Delta z = 1.5\times 10^{-2}$. Spectroscopic redshifts are also available for approximately $\sim 300,000$ entries, with an average error of $\Delta z = 1.5\times 10^{-4}$. Each galaxy's $B$ magnitude has its own estimated error, with an average of $\Delta B = 0.06^m$. We have replaced photometric redshifts of $\sim 40,000$ galaxies in 2MPZ with spectroscopic redshifts from the 2MASS Redshift Survey catalogue\footnote{\url{tdc-www.harvard.edu/2mrs/}} \citep{2012ApJS..199...26H}.

The SDSS-DR12Q catalogue contains $\sim$300,000 spectroscopically targeted and visually confirmed quasars from the Baryon Oscillation Spectroscopic Survey of the Sloan Digital Sky Survey III \citep{2013AJ....145...10D}. All of these quasars have photometric redshift values and corresponding errors associated to them.

We found it crucial to reduce the level of redundancy in our catalogue by eliminating duplicates. In order to achieve this goal, we applied a $k$-dimensional tree method \citep{361002.361007}, which builds a space-partitioning data structure, where a nearest neighbour search in parameter space can be done effectively with an $\mathcal{O}(\log N)$ time complexity. When matching two selected catalogues, we identified several duplicates by simply matching their object names. For all objects with common names, we calculated the differences between their RA, dec, $B$ magnitude, and $d_L$ values. After confirming empirically that all four sets of differences are normally distributed, we measured the corresponding standard deviations (see Table \ref{tab:params}), and applied a chi-squared test to the galaxies that are not part of the exactly matched-by-name set. The degrees of freedom in the chi-squared tests depended on the number of parameters from the set of four (RA, dec, $B$, and $d_L$) that the matched entries had in common. A threshold of 99 percent was used in the chi-squared test, which means that only 1 percent of duplicates were falsely identified as being two different objects. 

The advantage of this cross-matching method is that it can differentiate between galaxies residing in close proximity of each other by chance, but having significantly different redshifts and/or $B$ magnitudes. The number of accidental proximities of different galaxies rise with the number of entries in the merged catalogues, which makes it necessary to use more parameters beyond sky coordinates in the matching process. We first cross-matched GWGC with HyperLEDA, and 2MASS XSC with 2MPZ (where both catalogues use the same notation for the objects, so the duplicates have been identified simply by matching the object names), and then cross-matched the two resulting catalogues. As we found no objects matching between this catalogue and SDSS-DR12Q, we simply merged these two catalogues with each other.

\begin{table}
\caption{Standard deviations of differences between RA, dec, $B$ magnitude, and luminosity distance ($d_L$) values of pairs matching by their object names in two cross-matched catalogs (see Section \ref{sec:Compilation} for details). We give these standard deviations, as well as the number of such pairs ($N$) found in the two steps of (i) cross-matching GWGC with HyperLEDA (left column), and (ii) cross-matching [2MASS XSC + 2MPZ] with [GWGC + HyperLEDA] (right column).}
\label{tab:params}
\begin{tabular}{lcc}
\hline
 & GWGC $\times$  & [2MASS XSC + 2MPZ] $\times$\\
&  HyperLEDA & [GWGC + HyperLEDA]\\
\hline
$\sigma_{\mathrm{RA}}$ [deg] & $1.2 \times 10^{-3}$ & $3.0 \times 10^{-4}$\\
$\sigma_{\mathrm{dec}}$ [deg] & $7.9 \times 10^{-4}$ & $9.0 \times 10^{-5}$\\
$\sigma_{B}$ [mag] & 0.588 & 0.320\\
$\sigma_{d_L}$ [Mpc] & 29.1 & 93.5\\
$N$ & 28,279 & 152,894\\
\hline
\end{tabular}
\end{table}

\begin{figure*}
\includegraphics[width=2\columnwidth]{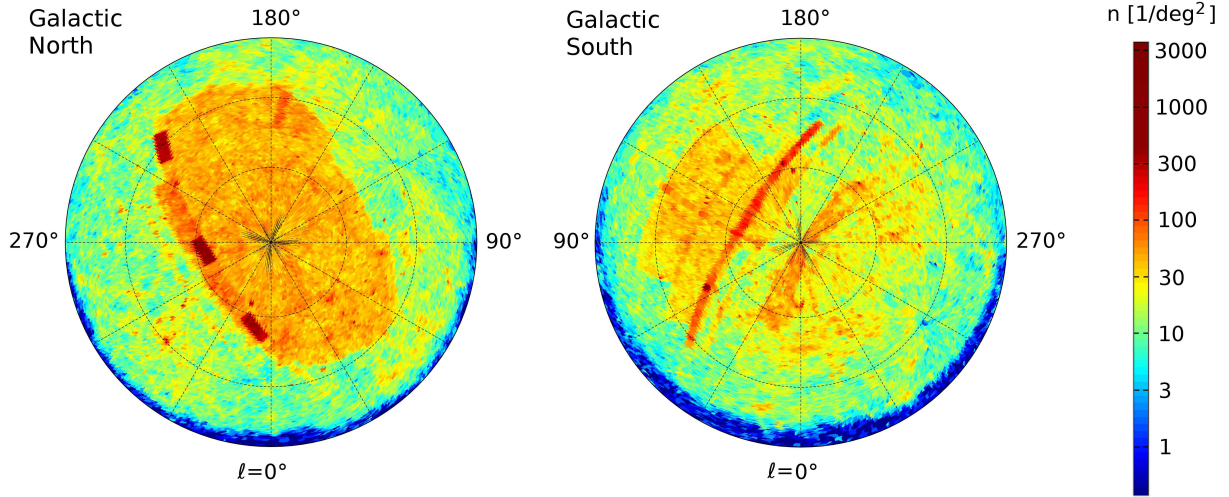}
\caption{The number density ($n$) of objects in GLADE, using azimuthal projection with galactic coordinates. The plane of the Milky Way obscures the visibility of background galaxies near the edges of the two plots. Overdense (red) patches and stripes, originating from the HyperLEDA catalogue (see \citealt{2014A&A...570A..13M}), show up as a result of deeper, more sensitive surveys, that have been made towards the corresponding sky directions.}
\label{fig:Mollweide}
\end{figure*}

Cross-matching the five catalogues with our method resulted with 2,965,718 galaxies, 297,014 quasars and 149 globular clusters (i.e. a total of 3,262,881 objects) in the GLADE catalogue. The sky distribution of all GLADE objects is shown in Figure \ref{fig:Mollweide} as a number density plot. The main reason of anisotropy in the distribution of GLADE objects is the different sensitivity of surveys and the uneven number of observations made toward different sky directions (note that the densest strips in the plot correspond to the HyperLEDA catalogue). The plane of the Milky Way is also noticeable, as the gas and dust content of the disk significantly reduces the visibility of galaxies in the background at optical and infrared wavelengths, resulting in a lower number density of GLADE objects along the galactic plane.

Redshifts of GLADE objects (where they are available) needed corrections for the objects' peculiar motions, in order to arrive at more accurate distance estimates from Hubble's law. We used the peculiar velocity field published in \cite{2015MNRAS.450..317C} for this correction, where peculiar velocities are defined at vertices of a uniform grid with 1.56 Mpc/$h$ linear resolution, within a distance limit of $\pm~200~\mathrm{Mpc}/h$ . We estimated the peculiar velocity of each GLADE object using a trilinear interpolation from peculiar velocity values at the eight nearest vertices, corrected the redshift of the object with the obtained value, and repeated the same process over and over again until the result converged to a certain redshift value (we found that all redshift values successfully converged).

GLADE as a final product is available as a $\mathtt{txt}$ file on the GLADE public website \footnote{GLADE website: \url{http://glade.elte.hu}}. Columns of each line of the file contain the following data (if available) for a single GLADE object:

1: Principal Galaxies Catalogue number

2: Name in the GWGC catalogue

3: Name in the HyperLEDA catalogue

4: Name in the 2MASS XSC catalogue

5: Name in the SDSS-DR12Q catalogue

6: Object type flag. Q: the source is from the SDSS-DR12Q catalogue. C: the source is a globular cluster. G: the source is not from the SDSS-DR12Q catalogue and not identified as a globular cluster

7: Right ascension in degrees

8: Declination in degrees

9: Luminosity distance in Mpc

10: Error of luminosity distance in Mpc

11: Redshift

12: Apparent $B$ magnitude

13: Absolute error of apparent $B$ magnitude

14: Absolute $B$ magnitude

15: Apparent $J$ magnitude

16: Absolute error of apparent $J$ magnitude

17: Apparent $H$ magnitude

18: Absolute error of apparent $H$ magnitude

19: Apparent $K$ magnitude

20: Absolute error of apparent $K$ magnitude

21: Luminosity distance measurement flag. $0$: the object has neither measured distance nor measured redshift value. $1$: the object has measured redshift value from which we have calculated its distance. $2$: the object has measured distance value from which we have calculated its redshift. $3$: we have replaced the measured photometric redshift with spectroscopic redshift from the 2MASS Redshift Survey catalogue, from which we have calculated its distance.

22: Velocity correction flag. It indicates whether the peculiar velocity correction was not ('0') or was ('1') applied when the distance of the object was calculated.

\section{Completeness}
\label{sec:Completeness}

We quantify the completeness of GLADE with two different methods: (i) by comparing, within different luminosity distance limits, the integrated $B$ luminosity of GLADE galaxies to calculated reference values (see details in next paragraphs), and (ii) by comparing luminosity distributions of GLADE galaxies within different luminosity distance shells to the Schechter function. In this paper, we only give completeness values up to $d_{\mathrm{L}}=200~\mathrm{Mpc}$. Since there are only 2 quasars in GLADE with $d_{\mathrm{L}}\leq 200$ Mpc, and only 5 percent of galaxies with $d_{\mathrm{L}}\leq 200~\mathrm{Mpc}$ not having $B$ magnitude data, we simply excluded quasars and such galaxies from our completeness measurements. 

The first method we applied is the one used in \citet{2011CQGra..28h5016W}, which allows a direct comparison between the completeness of GWGC and of GLADE. With this method, we compared the integrated $B$-band luminosity of galaxies in GLADE up to different distance limits to the same values expected for a complete catalogue of homogeneously distributed galaxies. According to \citet{2008ApJ...675.1459K}, such a complete sample of galaxies should have an average $B$-band luminosity density of \mbox{$(1.98 \pm 0.16) \times 10^{-2} \ L_{10} \ \mathrm{Mpc}^{-3}$,} where $L_{10}=10^{10} \ L_{B,\odot}$ and $L_{B,\odot}$ is the solar luminosity in the $B$-band, i.e. \mbox{$L_{B,\odot}=2.16 \times 10^{33} \ \mathrm{erg}/\mathrm{s}$}. The average $B$-band luminosity density was estimated by \cite{2003ApJ...592..819B} up to $z=0.1$ (i.e. $d_{\mathrm{L}}=420 \ \mathrm{Mpc}$), however \citet{2007ApJ...665..265F} has showed that the same value is valid within measurement errors up to $z=0.3$ (i.e. $d_{\mathrm{L}}=1.2 \ \mathrm{Gpc}$).

\begin{figure*}
\includegraphics[width=2\columnwidth]{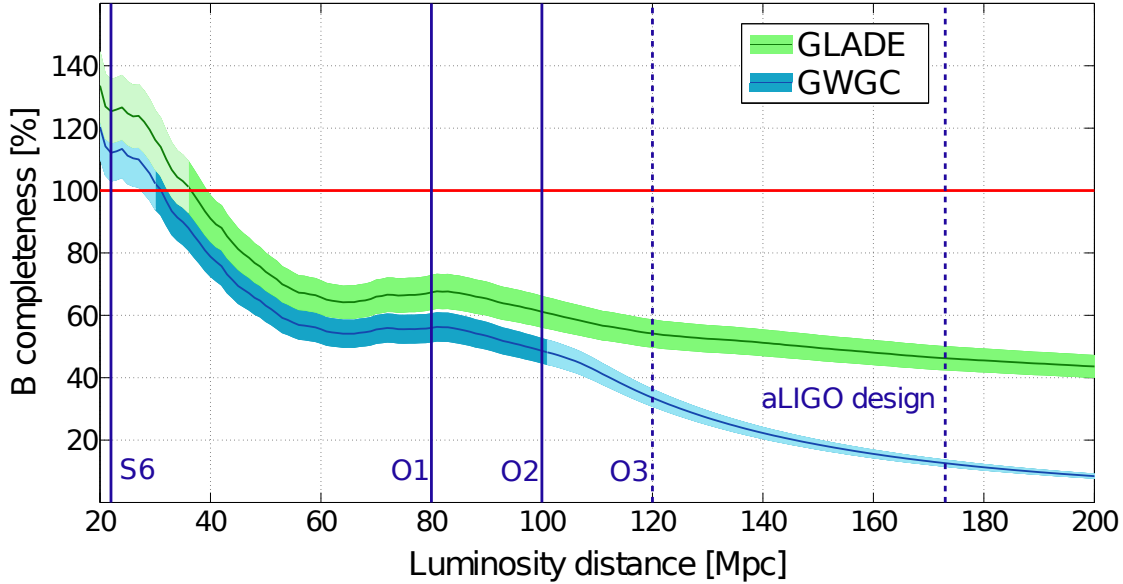}
\caption{The plots show the normalized integrated $B$-band luminosity of galaxies in GLADE (green) and in GWGC (blue) within luminosity distances indicated on the $x$-axis. Note that before producing this plot, we have updated all luminosity distances and $B$-band luminosities of GWGC galaxies where more recent measurements were available. The normalization is carried out with the integrated $B$-band luminosity calculated from the average $B$-band luminosity density of a complete catalogue of homogeneously distributed galaxies in the $z<0.1$ local universe (see \citealt{2008ApJ...675.1459K} for details). Faded segments of the curves (within 30 Mpc for GWGC, and within 37 Mpc for GLADE) indicate the fact that completeness values exceed 100 percent within these radii due to a local overdensity of galaxies around the Milky Way within these radii. The curve for GWGC is faded over 100 Mpc as well, since the catalogue applies a cutoff at this distance, which results in sharp decrease of its completeness (see Section \ref{sec:Compilation} for details). Vertical lines represent single-detector BNS ranges at various stages of development of LIGO detectors. The solid line marked with 'S6' shows the highest BNS range value that was achieved by H1 during the sixth science run of Initial LIGO \citep{2012arXiv1203.2674T}. The solid lines with the 'O1' and 'O2' labels represent the highest BNS ranges obtained by H1 during the O1 (80$~$Mpc) and O2 (100$~$Mpc) runs, respectively. The dashed lines marked with the labels 'O3' and 'aLIGO design' are the planned BNS ranges of a single aLIGO detector during O3 (at least 120$~$Mpc, see \citealt{2016LRR....19....1A}) and at design sensitivity (173$~$Mpc, see \citealt{Barsottietal}), respectively. Widths of the lighter green and blue stripes represent errors arising from uncertainties in the averaged $B$-band luminosity density value.} 
\label{fig:completeness}
\end{figure*}

In Figure \ref{fig:completeness} we show a comparison of the completeness of GLADE and GWGC based on our first completeness measure. Note that before producing this plot, we have updated all luminosity distances and $B$-band luminosities of GWGC objects where more recent measurements of these were available. As shown in Figure \ref{fig:completeness}, based on this completeness measure, GLADE is complete up to $\sim$37 Mpc, has a completeness of $\sim$61 percent within the maximal value of single-detector BNS ranges for aLIGO during O2 ($\sim$100 Mpc), $\sim$54 percent within the minimal planned BNS range during O3, and $\sim$48 percent within the planned BNS range of single aLIGO detectors with design sensitivity ($\sim173~\mathrm{Mpc}$, see \citealt{Barsottietal}).

\begin{figure*}
\includegraphics[width=2\columnwidth]{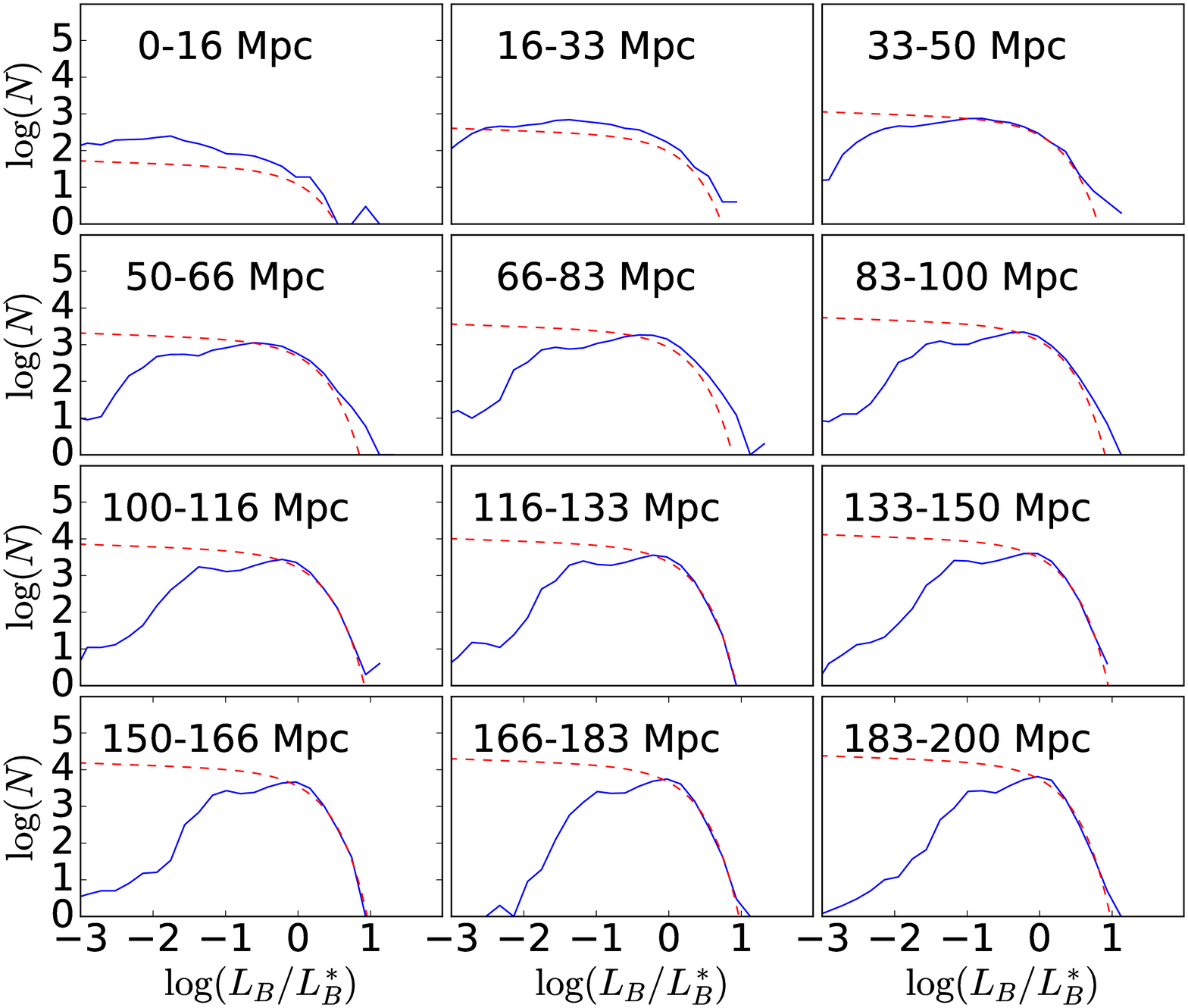}
\caption{Luminosity histograms of GLADE galaxies within different distance shells in terms of their measured $B$-band luminosities (blue solid line), compared to the same histograms we expect for complete catalogues based on $B$-band Schechter function measurements (red dashed line). We constructed this figure in a way to allow a direct comparison with the histograms shown in Figure 2 of \citet{2016ApJ...820..136G} for the Census of the Local Universe catalogue. $L_{B}^{*} = (1.2\,\pm\,0.1)\times10^{10} \ h^{-2} \ L_{B,\odot}$ is the characteristic luminosity of the Schechter function.}
\label{shells}
\end{figure*}

A second method was used to determine the completeness of GLADE using the Schechter function with parameters $\phi^{*} = (1.6\,\pm\,0.3)\times10^{-2} \ h^3 \ \mathrm{Mpc}^{-3}$, $a = -1.07\,\pm\,0.07$, $L_{B}^{*} = (1.2\,\pm\,0.1)\times10^{10} \ h^{-2} \ L_{B,\odot}$ \citep{2016ApJ...820..136G}. We divided galaxies into twelve luminosity distance shells, each having a width of $\Delta d_{\mathrm{L}}=16.7~\mathrm{Mpc}$. We constructed histograms of $B$-band luminosities in GLADE for each shell, which are plotted in Figure \ref{shells}, together with the corresponding Schechter functions. Figure \ref{shells} shows that as distance increases, more and more faint galaxies are missing from GLADE, while the histograms of $B$-band luminosities in GLADE exceed the Schechter functions in the first two bins due to a local overdensity of galaxies around the Milky Way.

\begin{figure}
\includegraphics[width=\columnwidth]{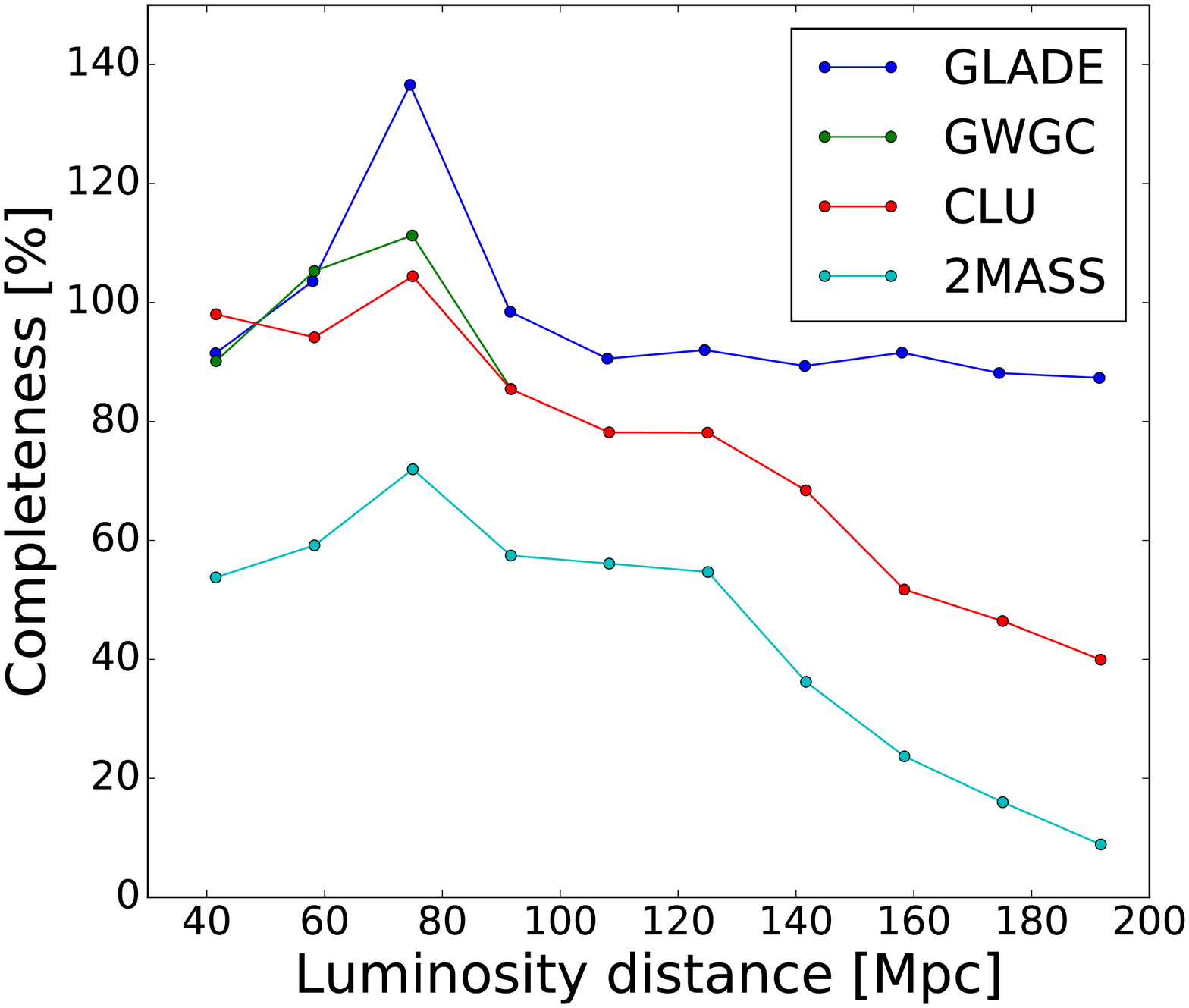}
\caption{Completeness in various distance bins (having a 16 Mpc width) of four different galaxy catalogues relative to the Schechter function for the brighter half of galaxies. Data for the GWGC, CLU, and 2MASS catalogues are taken directly from \citet{2016ApJ...820..136G}, and for GLADE it has been calculated using the same method. It can be seen that GLADE contains all the bright galaxies giving half of the total $B$ luminosity up to $d_L=91~$Mpc, and even at larger distances GLADE has a higher completeness than the other three catalogues. Note, that we used only the brighter half of all the galaxies for producing this figure, so even if a catalogue has 100 percent completeness, it can lack dimmer galaxies. Also note, that the peak visible in the 66-83 Mpc bin for all four completeness curves is due to the presence of the Perseus-Pisces Supercluster \citep{1981ApJ...243..411G} within this bin.}
\label{fig:shells2}
\end{figure}

The integration of the Schechter function by $x = L_{B}/L_{B}^{*}$ yields the following formula:
\begin{equation}
 \phi^{*}L_{B}^{*}\int_{x_1}^{\infty}x^{a+1}e^{-x}\mathrm{d}x = \phi^{*}L_{B}^{*}\Gamma(a+2,x_1),
\end{equation}
where following \cite{2016ApJ...820..136G}, we have chosen $x_1 = 0.626$, which (according to the Schechter function) is the integral bound that divides the galaxies into two subsets with equal total luminosities. We are interested in the brighter subset of these galaxies, as they would be the first targets of EM follow-up observations. In order to estimate the completeness of GLADE, we compared the integrated $B$-band luminosity of the brighter subset in each shell to the corresponding expected value calculated from the Schechter function. The completeness of GLADE in each shell is plotted in Figure \ref{fig:shells2}, together with the completeness values for GWGC, the 2MASS catalogue, and the CLU catalogue, from \citet{2016ApJ...820..136G}.

\section{Applications of GLADE}
\label{sec:Applications}

The purpose of GLADE is to (i) help identifications of host candidates for GW events, (ii) support target selections for EM follow-up observations of GW candidates, (iii) provide input data on the matter distribution of the local universe for astrophysical or cosmological simulations, and (iv) help identifications of host candidates for poorly localised EM transients, such as GRBs observed with the IPN. In this section we show how GLADE has already been used in these areas by several collaborations.

GLADE has been used in identifications of host candidates and in target selections for EM follow-up observations of GW events by multiple collaborations. Following the detection of the first GW signal from a coalescencing BNS (GW170817, see \citealt{2017PhRvL.119p1101A}) an extensive observing campaign was launched, which led to the identification of an EM counterpart by multiple observer partners of the LIGO-Virgo Collaboration \citep{2017ApJ...848L..12A}. Several of these teams used GLADE to maximize the chance of detecting the counterpart. For example, \cite{2017Natur.551...64A} found the optical counterpart with the Las Cumbres Observatory global network of robotic telescopes by targeting specific galaxies within the localisation area chosen from GLADE. \cite{Castro-Tirado} used the JGT robotic telescope at the BOOTES-5 station to image 15 galaxies from GLADE, resulting in finding the optical counterpart in the outskirts of \mbox{NGC 4993}. \cite{Gradoetal} used the GWsky tool to generate a pointing for the VLT Survey Telescope around the maximum probability pixel, however \mbox{NGC 4993} was not inside this pointing. GLADE is used as an input in GWsky, which is an interactive tool that tiles the localisation area of a GW event based on specific telescope parameters \citep{Grecoetal}\footnote{An additional web-based tool that uses GLADE as an input is Skymap Viewer, which interactively shows skymaps and host galaxy candidates for GW events (see \citealt{2016arXiv161100790L} and \url{losc.ligo.org/s/skymapViewer/}).}. Note that GWsky is used by the GRAWITA collaboration \citep{2017arXiv171005915B} as well. The Pi of the Sky robotic telescope surveyed an area that partially overlapped with the initial localisation area of GW170817, and \cite{Batschetal} searched for transients connected with objects from GLADE, however the telescope did not target NGC 4993, and thus the counterpart was not found. The HESS collaboration \citep{Naurois}, which operates the High Energy Stereoscopic System investigating cosmic gamma rays, carried out follow-up observations with three pointings derived from the initial LIGO skymap that was cross-correlated in three dimensions with GLADE. The first of these three pointings covered \mbox{NGC 4993}, which region they continued to monitor the following nights as well, however, their preliminary analysis did not reveal significant gamma-ray emission.

The multi-messenger observing of GW170817 allowed its use in measuring the local expansion rate of the universe, characterised by the Hubble constant (\citealt{2017Natur.551...85A}, for its theoretical background see \citealt{1986Natur.323..310S} and \citealt{2012PhRvD..86d3011D}). This was carried out by assuming NGC$~$4993 (the host of the EM counterpart, see \citealt{2017ApJ...848L..12A}) to be the host of the GW source producing GW170817. A similar analysis can be carried out using data only from the  GW observation itself. In such a calculation, a prior on the redshift can be obtained from galaxies inside the localisation volume, but it must be corrected for incompleteness of the galaxy sample. GLADE would be an ideal tool for this due to its high completeness and due to the fact that it directly contains all relevant galaxy data that is necessary for this type of an analysis.

We continuously use GLADE in identifying potential host galaxies for GRBs detected and poorly localised by the IPN. The identification process starts automatically whenever an IPN $3\sigma$ localisation error box for a GRB is published on the GCN Circular website\footnote{\url{gcn.gsfc.nasa.gov/gcn3_archive.html}}. The process identifies galaxies inside the error box, along with those residing in close proximity of it, to account for the possibility of a pair of neutron stars being kicked out from its host galaxy, and enter the error box before they merge and produce a short GRB. According to \citet{2013ApJ...776...18F}, 95\% of all BNSs kicked out from a galaxy should merge within a 100 kpc range around their hosts, therefore we decided to use this as our spatial offset limit. Our code shows the identified host candidates in the form of a table, and lists their luminosity distances, their angular distances from the closest point of the error box, their projected distances from the error box, and the probability that a BNS kicked out from the galaxy reaches the aforementioned projected distance before the merger. We calculated the probability using the observed distribution of projected physical offsets of short GRBs from their host galaxies, shown in \mbox{Fig. 5} of \cite{2013ApJ...776...18F}. Note, that in order to have a BNS coalescence inside the error box with this probability, one must assume that the kick occurred towards the closest point of the error box from the galaxy center. A skymap is also created showing the $\pm 0.2^{\circ}$ region around the error box with all GLADE galaxies indicated differently depending on the source of their distance data. Results of the analyses for different individual GRBs can be accessed directly on the GLADE website\footnote{\url{http://aquarius.elte.hu/glade/GRB.html}}.

Our analysis for GRB$~$150906B was of particular interest due to its proximity to a group of nearby galaxies identified as potential hosts \citep{2016arXiv161107947L}. The localised sky region for GRB$~$150906B \citep{Hurleyetal2015} lies close to NGC$~$3313 which has a luminosity distance of 54 Mpc, a distance that is well within the BNS range of a single aLIGO detector. Our pipeline has found one GLADE galaxy inside the error box and another one having an angular distance of 16.62'' and a projected distance of 40-52 kpc from the closest point of the GRB error box. The two galaxies have a luminosity distance of 536 $\pm$ 75 Mpc and 559 $\pm$ 75 Mpc, respectively. These distances are more compatible with predictions given by \citet{Ruffinietal2015} and \citet{Zhangetal2015} based on $E_{\mathrm{p}}-E_{\mathrm{iso}}$ and $E_{\mathrm{p}}-L_{\mathrm{iso}}$ relations for short GRBs ($z\sim 0.1$) than the galaxies of the NGC$~$3133 group, making the two galaxies far more probable hosts.

\section{Conclusions and Future Work}
\label{sec:Conclusions}

The GLADE catalogue is a value-added full-sky catalogue containing more than 3.26 million objects from which 2.97 million are categorized as galaxies, and the rest as quasars or globular clusters. It is complete up to $d_L=37^{+3}_{-4}~$Mpc in terms of cumulative $B$ luminosity of galaxies within distance $d_L$, and contains all of the bright galaxies giving half of the total $B$ luminosity up to $d_L=91~$Mpc. This high value of completeness and the presence of $B$ magnitudes and distances for most of the entries can make GLADE a useful tool for the identification of host candidates for GW events, for supporting target selections for EM follow-up observations of GW candidates, as well as for a handful of other purposes for the broader astronomical community. GLADE has already been used in several different projects, e.g. in finding the optical counterpart of GW170817, the first GW signal from a BNS coalescence, and in identifying potential host galaxies for poorly localised GRBs.

We are continuously improving GLADE in order to keep its suitability for EM follow-up efforts in the future. We are currently working on matching additional galaxy catalogues (such as WISE and Pan-STARRS) with GLADE in order to improve its completeness. Furthermore, we seek ways to develop the catalogue to include other relevant parameters, such as stellar mass and BNS formation rate estimates for the individual galaxies. Since the number density of GLADE objects is anisotropic (see Figure \ref{fig:Mollweide}), it may also be beneficial to provide completeness information constrained to different (or all) patches in the sky. We leave this task for future works on the framework of calculating catalogue completeness in 3D localisation volumes of individual GW events.

\section*{Acknowledgments}
This paper was reviewed by the LIGO Scientific Collaboration under LIGO Document P1800062.
The authors would like to thank Bence B\'{e}csy, Eric Chassande-Mottin, D\'{a}niel Erdei, Giuseppe Greco and \'{A}kos Sz\"{o}lgy\'{e}n for fruitful discussions throughout the project. The authors thank Christopher Berry, Maciej Bilicki and Erik Katsavounidis for their useful comments on the manuscript. This project has been supported by the Hungarian National Research, Development and Innovation Office -- NKFIH K-115709 and NN 114560. PR is supported through the \'UNKP-17-4 New National Excellence program of the Ministry of Human Capacities. The authors also gratefully acknowledge the Science and Technology Facilities Council of the United Kingdom. CM is supported by the Science and Technology Research Council (grant No. ST/ L000946/1). 
We are grateful for the Wide Field Astronomy Unit (WFAU) for providing the 2MPZ data used in creating GLADE.

\bsp

\label{lastpage}


\begin{thebibliography}{99}
\bibitem[Aab et al.(2016)]{2016PhRvD..94l2007A} Aab, A., Abreu, P., Aglietta, M., et al.\ 2016, \prd, 94, 122007 
\bibitem[Aasi et al.(2014a)]{2014ApJS..211....7A} Aasi, J., Abadie, J., Abbott, B.~P., et al.\ 2014a, \apjs, 211, 7 
\bibitem[Aasi et al.(2014b)]{2014PhRvL.113a1102A} Aasi, J., Abbott, B.~P., Abbott, R., et al.\ 2014b, Physical Review Letters, 113, 011102  
\bibitem[Aasi et al.(2015)]{2015CQGra..32g4001L} Aasi, J., Abbott, B.~P., et al.\ 2015, Classical and Quantum Gravity, 32, 074001
\bibitem[Abadie et al.(2012a)]{2012A&A...541A.155A} Abadie, J., Abbott, B.~P., Abbott, R., et al.\ 2012a, \aap, 541, A155  
\bibitem[Abadie et al.(2012b)]{2012A&A...539A.124L} Abadie, J., Abbott, B.~P., Abbott, R., et al.\ 2012b, \aap, 539, A124
\bibitem[Abadie et al.(2012c)]{2012arXiv1203.2674T} Abadie, J., Abbott, B.~P., Abbott, R., et al.\ 2012c, arXiv:1203.2674 
\bibitem[Abbott et al.(2016a)]{2016PhRvX...6d1015A} Abbott, B.~P., Abbott, R., Abbott, T.~D., et al.\ 2016a, Physical Review X, 6, 041015
\bibitem[Abbott et al.(2016b)]{2016LRR....19....1A} Abbott, B.~P., Abbott, R., Abbott, T.~D., et al.\ 2016b, Living Reviews in Relativity, 19, 1 
\bibitem[Abbott et al.(2016c)]{2016ApJ...826L..13A} Abbott, B.~P., Abbott, R., Abbott, T.~D., et al.\ 2016c, \apjl, 826, L13 
\bibitem[Abbott et al.(2016d)]{2016arXiv161107947L} Abbott, B.~P., Abbott, R., Abbott, T.~D., et al.\ 2016d, arXiv:1611.07947  
\bibitem[Abbott et al.(2017a)]{2017PhRvL.119n1101A} Abbott, B.~P., Abbott, R., Abbott, T.~D., et al.\ 2017, Physical Review Letters, 119, 141101 
\bibitem[Abbott et al.(2017b)]{2017ApJ...848L..12A} Abbott, B.~P., Abbott, R., Abbott, T.~D., et al.\ 2017a, \apjl, 848, L12 
\bibitem[Abbott et al.(2017c)]{2017PhRvL.119p1101A} Abbott, B.~P., Abbott, R., Abbott, T.~D., et al.\ 2017b, Physical Review Letters, 119, 161101 
\bibitem[Abbott et al.(2017d)]{2017Natur.551...85A} Abbott, B.~P., Abbott, R., Abbott, T.~D., et al.\ 2017, \nat, 551, 85 
\bibitem[Acernese et al.(2015)]{2015CQGra..32b4001A} Acernese, F., Agathos, M., Agatsuma, K., et al.\ 2015, Classical and Quantum Gravity, 32, 024001 
\bibitem[Arcavi et al.(2017)]{2017Natur.551...64A} Arcavi, I., Hosseinzadeh, G., Howell, D.~A., et al.\ 2017, \nat, 551, 64 
\bibitem[Baret et al.(2012)]{2012PhRvD..85j3004B} Baret, B., Bartos, I., Bouhou, B., et al.\ 2012, \prd, 85, 103004 
\bibitem[Barsotti et al.(2018)]{Barsottietal} Barsotti, L., McCuller, L., Evans, M., Fritschel, P.\ 2018, LIGO-T1800044-v4 Technical Note
\bibitem[Bartos et al.(2015)]{2015ApJ...801L...1B} Bartos, I., Crotts, A.~P.~S., \& M{\'a}rka, S.\ 2015, \apjl, 801, L1 
\bibitem[Batsch et al.(2017)]{Batschetal} Batsch, Castro-Tirado, A.~J., Czyrkowski, H.\ 2017, GCN Circular 21931
\bibitem[B{\'e}csy et al.(2017)]{2017ApJ...839...15B} B{\'e}csy, B., Raffai, P., Cornish, N.~J., et al.\ 2017, \apj, 839, 15 
\bibitem[Bell et al.(2003)]{2003ApJS..149..289B} Bell, E.~F., McIntosh, D.~H., Katz, N., \& Weinberg, M.~D.\ 2003, \apjs, 149, 289 
\bibitem[Bentley(1975)]{361002.361007} Bentley, J.~L.\ 1975, Commun. ACM, 18, 9
\bibitem[Berry et al.(2015)]{2015ApJ...804..114B} Berry, C.~P.~L., Mandel, I., Middleton, H., et al.\ 2015, \apj, 804, 114 
\bibitem[Bilicki et al.(2014)]{2014ApJS..210....9B} Bilicki, M., Jarrett, T.~H., Peacock, J.~A., Cluver, M.~E., \& Steward, L.\ 2014, \apjs, 210, 9 
\bibitem[Blanton et al.(2003)]{2003ApJ...592..819B} Blanton, M.~R., Hogg, D.~W., Bahcall, N.~A., et al.\ 2003, \apj, 592, 819 
\bibitem[Brocato et al.(2017)]{2017arXiv171005915B} Brocato, E., Branchesi, M., Cappellaro, E., et al.\ 2017, arXiv:1710.05915 
\bibitem[Carrick et al.(2015)]{2015MNRAS.450..317C} Carrick, J., Turnbull, S.~J., Lavaux, G., \& Hudson, M.~J.\ 2015, \mnras, 450, 317 
\bibitem[Castro-Tirado et al.(2017)]{Castro-Tirado} Castro-Tirado A.~J., Tello J.~C., Hu Y. et al.\ 2017, GCN 21624
\bibitem[Chassande-Mottin et al.(2011)]{2011GReGr..43..437C} Chassande-Mottin, E., Hendry, M., Sutton, P.~J., \& M{\'a}rka, S.\ 2011, General Relativity and Gravitation, 43, 437 
\bibitem[Coughlin et al.(2018)]{2018arXiv180302255C} Coughlin, M.~W., Tao, D., Chan, M.~L., et al.\ 2018, arXiv:1803.02255 
\bibitem[Cowperthwaite et al.(2016)]{2016ApJ...826L..29C} Cowperthwaite, P.~S., Berger, E., Soares-Santos, M., et al.\ 2016, \apjl, 826, L29 
\bibitem[Dawson et al.(2013)]{2013AJ....145...10D} Dawson, K.~S., Schlegel, D.~J., Ahn, C.~P., et al.\ 2013, \aj, 145, 10 
\bibitem[Del Pozzo(2012)]{2012PhRvD..86d3011D} Del Pozzo, W.\ 2012, \prd, 86, 043011 
\bibitem[Del Pozzo et al.(2018)]{2018arXiv180108009D} Del Pozzo, W., Berry, C., Ghosh, A., Haines, T., \& Vecchio, A.\ 2018, arXiv:1801.08009 
\bibitem[Essick et al.(2015)]{2015ApJ...800...81E} Essick, R., Vitale, S., Katsavounidis, E., Vedovato, G., \& Klimenko, S.\ 2015, \apj, 800, 81 
\bibitem[Evans et al.(2016)]{2016MNRAS.455.1522E} Evans, P.~A., Osborne, J.~P., Kennea, J.~A., et al.\ 2016, \mnras, 455, 1522  
\bibitem[\protect\citeauthoryear{Faber et al.}{2007}]{2007ApJ...665..265F} Faber, S.~M., Willmer, C.~N.~A., Wolf, C., et al.\ 2007, \apj, 665, 265 
\bibitem[Fan et al.(2014)]{2014ApJ...795...43F} Fan, X., Messenger, C., \& Heng, I.~S.\ 2014, \apj, 795, 43 
\bibitem[Finn \& Chernoff(1993)]{1993PhRvD..47.2198F} Finn, L.~S., \& Chernoff, D.~F.\ 1993, \prd, 47, 2198 
\bibitem[Fong et al.(2013)]{2013ApJ...769...56F} Fong, W., Berger, E., Chornock, R., et al.\ 2013, \apj, 769, 56 
\bibitem[Fong \& Berger(2013)]{2013ApJ...776...18F} Fong, W., \& Berger, E.\ 2013, \apj, 776, 18 
\bibitem[Gehrels et al.(2016)]{2016ApJ...820..136G} Gehrels, N., Cannizzo, J.~K., Kanner, J., et al.\ 2016, \apj, 820, 136 
\bibitem[Grado et al.(2017)]{Gradoetal} Grado, A., Cappellaro, E., Greco, G., et al.\ 2017, GCN Circular 21598
\bibitem[Greco et al.(2018 in prep.)]{Grecoetal} Greco, G., Branchesi, M., Stratta, G., et al., in prep.
\bibitem[Gregory et al.(1981)]{1981ApJ...243..411G} Gregory, S.~A., Thompson, L.~A., \& Tifft, W.~G.\ 1981, \apj, 243, 411 
\bibitem[Hambly et al.(2001)]{2001MNRAS.326.1279H} Hambly, N.~C., MacGillivray, H.~T., Read, M.~A., et al.\ 2001, \mnras, 326, 1279 
\bibitem[Hanna et al.(2014)]{2014ApJ...784....8H} Hanna, C., Mandel, I., \& Vousden, W.\ 2014, \apj, 784, 8
\bibitem[Huchra et al.(2012)]{2012ApJS..199...26H} Huchra, J.~P., Macri, L.~M., Masters, K.~L., et al.\ 2012, \apjs, 199, 26 
\bibitem[Hurley et al.(2013)]{2013EAS....61..459H} Hurley, K., Mitrofanov, I.~G., Golovin, D., et al.\ 2013, EAS Publications Series, 61, 459 
\bibitem[\protect\citeauthoryear{Hurley et al.}{2015}]{Hurleyetal2015} Hurley K. et al., 2015, GCN Circular 18258
\bibitem[Iyer at al.(2011)]{LIGO_India} Iyer, B., Souradeep, T., Unnikrishnan, C.~S., et al.\ 2011, LIGO Document M1100296
\bibitem[Jarrett et al.(2000)]{2000AJ....119.2498J} Jarrett, T.~H., Chester, T., Cutri, R., et al.\ 2000, \aj, 119, 2498 
\bibitem[Karachentsev et al.(2004)]{2004AJ....127.2031K} Karachentsev, I.~D., Karachentseva, V.~E., Huchtmeier, W.~K., \& Makarov, D.~I.\ 2004, \aj, 127, 2031 
\bibitem[Kasen et al.(2013)]{2013ApJ...774...25K} Kasen, D., Badnell, N.~R., \& Barnes, J.\ 2013, \apj, 774, 25 
\bibitem[\protect\citeauthoryear{Kopparapu et al.}{2008}]{2008ApJ...675.1459K} Kopparapu, R.~K., Hanna, C., Kalogera, V., et al.\ 2008, \apj, 675, 1459 
\bibitem[Li \& Paczy{\'n}ski(1998)]{1998ApJ...507L..59L} Li, L.-X., \& Paczy{\'n}ski, B.\ 1998, \apjl, 507, L59 
\bibitem[Li \& Williams(2016)]{2016arXiv161100790L} Li, K., \& Williams, R.\ 2016, arXiv:1611.00790 
\bibitem[Makarov et al.(2014)]{2014A&A...570A..13M} Makarov, D., Prugniel, P., Terekhova, N., Courtois, H., \& Vauglin, I.\ 2014, \aap, 570, A13 
\bibitem[Metzger et al.(2010)]{2010MNRAS.406.2650M} Metzger, B.~D., Mart{\'{\i}}nez-Pinedo, G., Darbha, S., et al.\ 2010, \mnras, 406, 2650 
\bibitem[de Naurois(2017)]{Naurois} de Naurois, M.\ 2017, GCN Circular 21674
\bibitem[Ochsenbein et al.(2000)]{2000A&AS..143...23O} Ochsenbein, F., Bauer, P., \& Marcout, J.\ 2000, \aaps, 143, 23 
\bibitem[P{\^a}ris et al.(2016)]{Paris} P{\^a}ris, I., Petitjean, P., Ross, N.~P., et al.\ 2016, arXiv:1608.06483
\bibitem[Paturel et al.(1988)]{1988ESOC...28..435P} Paturel, G., Bottinelli, L., Fouque, P., \& Gouguenheim, L.\ 1988, European Southern Observatory Conference and Workshop Proceedings, 28, 435 
\bibitem[Phinney(1991)]{1991ApJ...380L..17P} Phinney, E.~S.\ 1991, \apjl, 380, L17 
\bibitem[Prugniel \& Simien(1996)]{1996A&A...309..749P} Prugniel, P., \& Simien, F.\ 1996, \aap, 309, 749 
\bibitem[\protect\citeauthoryear{Ruffini et al.}{2015}]{Ruffinietal2015} Ruffini R. et al., 2015, GCN Circular 18296 
\bibitem[Schutz(1986)]{1986Natur.323..310S} Schutz, B.~F.\ 1986, \nat, 323, 310 
\bibitem[Singer et al.(2014)]{2014ApJ...795..105S} Singer, L.~P., Price, L.~R., Farr, B., et al.\ 2014, \apj, 795, 105  
\bibitem[Singer et al.(2016)]{2016ApJ...829L..15S} Singer, L.~P., Chen, H.-Y., Holz, D.~E., et al.\ 2016, \apjl, 829, L15 
\bibitem[Skrutskie et al.(2006)]{2006AJ....131.1163S} Skrutskie, M.~F., Cutri, R.~M., Stiening, R., et al.\ 2006, \aj, 131, 1163 
\bibitem[Somiya(2012)]{2012CQGra..29l4007S} Somiya, K.\ 2012, Classical and Quantum Gravity, 29, 124007 
\bibitem[Soneira \& Peebles(1978)]{1978AJ.....83..845S} Soneira, R.~M., \& Peebles, P.~J.~E.\ 1978, \aj, 83, 845 
\bibitem[Tanvir et al.(2013)]{2013Natur.500..547T} Tanvir, N.~R., Levan, A.~J., Fruchter, A.~S., et al.\ 2013, \nat, 500, 547 
\bibitem[Tully(1987)]{1987ApJ...321..280T} Tully, R.~B.\ 1987, \apj, 321, 280 
\bibitem[Tully et al.(2009)]{2009AJ....138..323T} Tully, R.~B., Rizzi, L., Shaya, E.~J., et al.\ 2009, \aj, 138, 323 
\bibitem[White et al.(2011)]{2011CQGra..28h5016W} White, D.~J., Daw, E.~J., \& Dhillon, V.~S.\ 2011, Classical and Quantum Gravity, 28, 085016 
\bibitem[Wright et al.(2010)]{2010AJ....140.1868W} Wright, E.~L., Eisenhardt, P.~R.~M., Mainzer, A.~K., et al.\ 2010, \aj, 140, 1868-1881 
\bibitem[\protect\citeauthoryear{Zhang, Zhang \& Zhang}{2015}]{Zhangetal2015} Zhang F., Zhang B., \& Zhang B., 2015, GCN Circular 18298

\end{thebibliography}
\end{document}